\documentclass[iop]{emulateapj}
\usepackage{apjfonts}
\usepackage{amsmath,amssymb,bm}

\renewcommand{\vec}[1]{\mathbf{#1}}

\newcommand{\diffp}[2]{\frac{\partial #1}{\partial #2}}

\shorttitle{Parallel Alfv\'enic Ion-Beam Instability}
\shortauthors{D.~Verscharen et al.}

\begin{document}

\title{A Parallel-Propagating Alfv\'enic Ion-Beam Instability in the High-Beta Solar Wind}


\author{Daniel Verscharen, Sofiane Bourouaine, and Benjamin D.~G.~Chandran \altaffilmark{1}}
\affil{Space Science Center, University of New Hampshire, Durham, NH 03824, USA; daniel.verscharen@unh.edu, s. bourouaine@unh.edu, benjamin.chandran@unh.edu}
\altaffiltext{1}{Also at Department of Physics, University of New Hampshire, Durham, NH 03824, USA}
\author{Bennett A.~Maruca}
\affil{Space Science Laboratory, University of California, Berkeley, CA 94720, USA; bmaruca@ssl.berkeley.edu}

\journalinfo{The Astrophysical Journal, 773:8 (8pp), 2013 August 10 }
\submitted{Received 2012 December 20; accepted 2013 June 11; published 2013 July 18}

\begin{abstract}
We investigate the conditions under which parallel-propagating
Alfv\'en/ion-cyclotron waves are driven unstable by an isotropic
($T_{\perp \alpha } = T_{\parallel\alpha}$) population of alpha particles drifting
parallel to the magnetic field at an
average speed $U_{\alpha}$ with respect to the protons. We derive an approximate analytic
condition for the minimum value of $U_{\alpha}$ needed to excite this
instability and refine this result using numerical solutions to the
hot-plasma dispersion relation. When the alpha-particle number density
is $\simeq 5\%$ of the proton number density and the two species have
similar thermal speeds, the instability requires that $\beta_{\rm p}
\gtrsim 1$, where $\beta_{\rm p}$ is the ratio of the proton pressure
to the magnetic pressure. For $1\lesssim \beta_{\mathrm p }\lesssim
12$, the minimum $U_{\alpha}$ needed to excite this instability ranges
from $0.7v_{\mathrm A}$ to $0.9v_{\mathrm A}$, where $v_{\mathrm A}$
is the Alfv\'en speed. This threshold is smaller than the threshold of
$\simeq 1.2v_{\mathrm A}$ for the parallel magnetosonic instability,
which was previously thought to have the lowest threshold of the
alpha-particle beam instabilities at $\beta_{\mathrm p}\gtrsim
0.5$. We discuss the role of the parallel Alfv\'enic drift instability for
the evolution of the alpha-particle drift speed in the solar wind. We also analyze measurements from the \emph{Wind} spacecraft's Faraday cups and show that the $U_{\alpha}$ values measured in solar-wind streams with $T_{\perp \alpha }\approx T_{\parallel\alpha}$ are approximately bounded from above by the threshold of the parallel Alfv\'enic instability.
\end{abstract}


\section{Introduction}

The solar wind is a plasma consisting of electrons, protons, and other
ion species. Among those ions, the alpha particles play a particularly important
role for the dynamics and thermodynamics of the solar wind since their
mass density is typically about 15\% - 20\%  of the proton mass
density. It has been known for a long time that the alpha particles in
the fast solar wind at $0.3\,\mathrm{AU}<r<4.2\,\mathrm{AU}$ drift
with respect to the protons with a typical speed of order the local
proton Alfv\'en speed
\begin{equation}
v_{\mathrm A}\equiv \frac{B}{\sqrt{4 \pi n_{\mathrm p}m_{\mathrm p}}},
\end{equation}
where $n_{\mathrm p}$ and $m_{\mathrm p}$ are the proton number
density and the proton mass, respectively, and $B$ is the magnetic
field strength \citep{marsch82,marsch87,reisenfeld01}. Since
$v_{\mathrm A}$ decreases with increasing heliocentric distance $r$
outside the corona, and since the proton outflow velocity varies only
weakly with~$r$ for $r>0.3 \mbox{ AU}$, the observed
limitation of the alpha particle drift reflects a continuous
deceleration of the alpha particles. Micro-instabilities driven by the
relative drift are believed to be responsible for this deceleration
\citep{isenberg83,marsch87,gomberoff96,gary00,goldstein00,gary03,verscharen13}. 

We define the quantity
\begin{equation}
\beta_j\equiv\frac{8\pi n_j k_{\mathrm B}T_j}{ B^2},
\end{equation}
where $n_j$ and $T_j$ are the number density and temperature of species $j$. Spacecraft measurements show a large variety and (on average) a radial increase of $\beta_j$ with increasing $r$ over the range $0.3\,\mathrm{AU}<r<1\,\mathrm{AU}$ probed by the \emph{Helios} satellites \citep{marsch82,marsch82b}. In the fast solar wind, the typical value of $\beta_{\mathrm p}$ at 1 AU is of order unity, and the mean free path for particle collisions is of the same order as the distance from the Sun, which indicates that kinetic effects play an important role and should be included in a complete description of the plasma.

Additional important factors for the behavior of waves and
instabilities in plasmas are the temperature ratios between the
species.  The measured temperatures at 1~AU indicate that in slow-solar-wind streams collisional relaxation can lead to equal proton and
alpha temperatures~\citep{kasper08}. In the less collisional fast
wind, the distribution of $T_{\alpha}/T_{\mathrm p}$ values peaks at
$T_{\alpha}/T_{\mathrm p}\approx 4$, consistent with roughly equal
thermal speeds of protons and alpha particles
\citep{marsch82,kasper08}. Possible explanations for the enhanced
alpha-particle temperatures include cyclotron heating
\citep{marsch82c,isenberg83,vonsteiger95,neugebauer96,reisenfeld01}, stochastic heating
\citep{chandran10}, and transit-time damping \citep{lynn12}.

The properties of plasma waves in a high-$\beta_{\mathrm p}$ environment can significantly differ from their properties in the cold-plasma limit \citep{gary86}. The dispersion relations of these waves change, and resonant damping plays a more and more important role with increasing temperature. The thresholds of beam instabilities also depend on $\beta_j$, and some instabilities are only active in a small range of $\beta_j$ values \citep[e.g., ][]{montgomery76,li00}. 
The previous literature on ion drifts in the solar wind describes two Alfv\'enic instabilities that propagate obliquely to the background magnetic field and operate at low $\beta_{\mathrm p}$ and $\beta_{\alpha}$ with threshold speeds less than $1.3v_{\mathrm A}$ only at $\beta_{\alpha}<0.1$. Within this literature, the only instability that acts at higher $\beta_{\mathrm p}$ and $\beta_{\alpha}$ is the magnetosonic instability. Its  growth rate is highest at parallel propagation. However, unless temperature anisotropies with $T_{\perp}<T_{\parallel}$ are present, it requires drift speeds $\gtrsim 1.2v_{\mathrm A}$ to become unstable \citep{gary00b}. In contrast, the observed differential flows are generally smaller than $v_{\mathrm A}$ in both the slow wind and fast wind  \citep{marsch82,reisenfeld01}, which indicates that the parallel magnetosonic instability is not excited in solar-wind streams near 1 AU in which $T_{\perp \alpha}\simeq T_{\parallel \alpha}$.

An instability of the parallel-propagating Alfv\'en/ion-cyclotron wave in the presence of a hot beam has been discussed in the literature before. \citet{gary93} describes this ion/ion left-hand resonant instability in his Figures 8.1 and 8.3 for a plasma consisting of one ion species. 
In this paper, we present new results on a parallel-propagating Alfv\'enic drift
instability that is excited by drifting alpha particles with $T_{\perp \alpha}= T_{\parallel\alpha}$. The threshold of this drift speed is between $0.7v_{\mathrm A}$
and $0.9v_{\mathrm A}$, depending on the values of $\beta_{\alpha}$ and $\beta_{\mathrm p}$. We describe the local effects of this instability on the alpha particles and discuss the waves that are generated once the instability threshold is exceeded. We do not undertake the more ambitious task of developing a complete picture of the alpha-particle evolution in the solar wind, which would require us to include additional effects such as the interplay of different instabilities, collisions, and local heating. In Section~\ref{dispersions}, we discuss the dispersion relation of
parallel Alfv\'en/ion-cyclotron waves in the cold-plasma approximation. We also discuss the hot-plasma dispersion relation and a
numerical code that we have developed to solve this dispersion
relation. In Section~\ref{sec:QLT} we review some general properties of
resonant wave--particle interactions, and in Section~\ref{sect_cold} we
derive an approximate analytic expression for the instability
threshold. In Section~\ref{hot_plasma}, we use the full dispersion
relation of the hot plasma to test these analytical results and
quantify the growth rate of this instability.  While we principally focus on isotropic temperature, we also consider  the effects
of temperature anisotropies in Section~\ref{hot_plasma}. We describe the
possible quasilinear evolution of the alpha-particle distribution
function in the presence of this instability in
Section~\ref{sec:comp}. In Section \ref{sect:obs} we show that the theoretical instability threshold provides an approximate upper bound to the $U_{\alpha}$ values in solar-wind streams with $T_{\perp \alpha}\simeq T_{\parallel\alpha}$ as measured by the \emph{Wind} spacecraft at 1 AU. In Section~\ref{conclusions} we summarize our
results and discuss the relevance of this instability to the evolution
of alpha particles in the fast solar wind.

\section{The Cold-Plasma and Hot-Plasma Dispersion Relations}\label{dispersions}

We limit ourselves to wavevectors that are parallel to the background magnetic field $\vec B_0=B_0\hat{\vec e}_z$. In the cold-plasma approximation, the dispersion relation for the Alfv\'en/ion-cyclotron-wave solutions is \citep{gomberoff91}
\begin{equation}\label{colddisp}
\frac{k_{\parallel}^2v_{\mathrm A}^2}{\Omega_{\mathrm p}}=\frac{\omega^2}{\Omega_{\mathrm p}-\omega}+\frac{4\eta(\omega-k_{\parallel}U_{\alpha})^2}{\Omega_{\mathrm p}-2\omega+2k_{\parallel}U_{\alpha}},
\end{equation}
 where $k_{\parallel}$ is the field-parallel component of the wavevector, $\omega=\omega_{\mathrm r}+i\gamma$ is the complex wave frequency, $\Omega_{\mathrm p}\equiv q_{\mathrm p}B_0/m_{\mathrm p}c$ is the proton cyclotron frequency, $\eta\equiv n_{\alpha}/n_{\mathrm p}$ is the fractional alpha-particle density, and $U_{\alpha}$ is the drift speed of the alpha particles with respect to the protons.  Equation~(\ref{colddisp}) is based on the zero-space-charge and zero-current conditions,
\begin{align}
n_{\mathrm e}&=\frac{1}{e}\sum\limits_i n_{i}q_{i}, \label{zerocharge} \\
U_{\mathrm e}&=\frac{1}{n_{\mathrm e} e}\sum\limits_i n_{i}q_iU_i,\label{zerocurrent}
\end{align}
where the sum is taken over all ion species $i$, which in our case
means protons and alpha particles. Since we work in the proton frame,
$U_{\mathrm p}=0$.  Both ion species are associated with an
ion-cyclotron-wave branch in the solutions of the dispersion
relation. We will focus on the so called Alfv\'en/proton-cyclotron
(A/PC) branch only, which is the solution for which $\omega_{\mathrm
  r}\rightarrow \Omega_{\mathrm p}$ as $k_{\parallel}\rightarrow
+\infty$.

In the more general case of a hot plasma (i.e.,~a plasma with nonzero temperature), the linear dispersion relation is based on Maxwell's equations along with the Vlasov equation, 
\begin{equation}
\diffp{f_j}{t}+\vec v \cdot \diffp{f_j}{\vec x}+\frac{q_j}{m_j}\left(\vec E+\frac{1}{c}\vec v\times \vec B \right)\cdot \diffp{f_j}{\vec v}=0
\end{equation}
for the particle species $j$ with charge $q_j$ and mass $m_j$ in an
electric field $\vec E$ and a magnetic field $\vec B$ \citep{stix92}. The distribution function $f_j$ is
written as $f_j=f_{0j}+\delta f_j$ with a homogeneous,
time-independent background $f_{0j}$ and a small perturbation $\delta
f_j$. The Vlasov and Maxwell's equations are then linearized to
describe the evolution of $\delta f_j$ and the fluctuating
electromagnetic fields.  We assume that each species' background
distribution function can be approximated as a drifting bi-Maxwellian
in cylindrical coordinates in $v$-space:
\begin{equation}
f_{0j}=\frac{n_j}{\pi^{3/2}w_{\perp j }^2w_{\parallel j }}\exp\left(-\frac{v_{\perp}^2}{w_{\perp j }^2}-\frac{\left(v_{\parallel}-U_j\right)^2}{w_{\parallel j }^2}\right),
\end{equation}
where
\begin{equation}
w_{\perp j }\equiv \sqrt{\frac{2k_{\mathrm B}T_{\perp j}}{m_j}}\quad\text{and}\quad
w_{\parallel j}\equiv \sqrt{\frac{2k_{\mathrm B}T_{\parallel j}}{m_j}}
\end{equation}
are the perpendicular and parallel thermal speeds, and $v_\perp$ and
$v_\parallel$ are the components of~$\vec{v}$ perpendicular and
parallel to the background magnetic field~$\vec{B}_0 = B_0 \hat{\vec e}_z$.
A long but straightforward calculation \citep[Chapt.~10
  from][]{stix92} eventually allows one to calculate the dielectric
tensor $\varepsilon$. The dispersion relation is then given by
\begin{equation}\label{fulldisp}
\frac{\vec kc}{\omega}\times\left( \frac{\vec kc}{\omega}\times \vec E_k\right)+\varepsilon \vec E_{k}\equiv \mathcal D\vec E_k=0.
\end{equation}

To solve this dispersion relation numerically, we developed a code
named NHDS (New Hampshire Dispersion relation Solver). The linearized
Vlasov-Maxwell system is solved for arbitrary directions of
propagation with respect to the background field by a secant method
allowing for an arbitrary number of particle species with given
charge, mass, temperatures, temperature anisotropies, densities, and
drift speeds. NHDS evaluates all calculations in double precision. For the general case when the angle between $\vec k$ and $\vec B_0$ is nonzero, the code
sums (over index~$n$) the modified Bessel functions $I_n(\lambda_j)$ that occur
in Equation~(\ref{fulldisp}) until $I_n(\lambda_j) < 10^{-50}$, where $\lambda_j\equiv k_{\perp}^2w_{\perp j}^2/2\Omega_j^2$. An initial guess for $\vec k$ and $\omega$
must be provided as the starting point of the secant calculation. This
initial guess defines the mode that is then tracked by the code as it
scans through different values of $\vec k$.  The electron density
$n_{\mathrm e}$ and electron drift speed $U_{\mathrm e}$ are again
adjusted according to the zero-space-charge and zero-current
conditions in Equations~(\ref{zerocharge}) and (\ref{zerocurrent}). NHDS
solutions have been benchmarked against the literature on
micro-instabilities such as the ion-cyclotron, firehose, mirror mode,
ion-ion, and electron-heat-flux instabilities \citep{gary93} and the
known dispersion relations of ion-cyclotron, whistler, lower-hybrid,
ion-Bernstein, and kinetic Alfv\'en waves.

\section{Resonant Wave--Particle Interactions}
\label{sec:QLT} 

In the limit of small wave amplitudes and small growth/damping rates, the evolution of the distribution function is described by quasilinear theory. Resonant particles undergo a diffusion process in velocity space according to the equation
\begin{multline}
\diffp{f_j}{t}=\lim _{V\to \infty}\sum \limits_{n=-\infty}^{+\infty}\frac{q_j^2}{8 \pi^2m_j^2}\int \frac{1}{Vv_{\perp}}\hat Gv_{\perp}\delta(\omega_{k\mathrm r}-k_{\parallel}v_{\parallel}-n\Omega_j)\\
\times \left|\psi_{n,k} \right|^2\hat Gf_j\mathrm d^3k,\label{qldiff}
\end{multline}
where
\begin{equation}
\hat G\equiv\left(1-\frac{k_{\parallel}v_{\parallel}}{\omega_{k\mathrm r}}\right)\diffp{}{v_{\perp}}+\frac{k_{\parallel}v_{\perp}}{\omega_{k\mathrm r}}\diffp{}{v_{\parallel}}
\end{equation}
and 
\begin{multline}
\psi_{n,k}\equiv \frac{1}{\sqrt{2}}\left[E_{k,\mathrm r}e^{i\phi}J_{n+1}(\sigma_j)+E_{k,\mathrm l}e^{-i\phi}J_{n-1}(\sigma_j)\right]\\
+\frac{v_{\parallel}}{v_{\perp}}E_{kz}J_n(\sigma_j)
\label{eq:psink} 
\end{multline}
\citep{kennel66,marsch06}. The Fourier-transformed electric-field
vector $(E_{kx}, E_{ky}, E_{kz})$ is used to define the quantities
$E_{k,\mathrm r}\equiv(E_{kx}-iE_{ky})/\sqrt{2}$ and $E_{k,\mathrm
  l}\equiv(E_{kx}+iE_{ky})/\sqrt{2}$. The argument of the $n$th order Bessel
function $J_n$ is defined as $\sigma_j\equiv
k_{\perp}v_{\perp}/\Omega_j$, and the azimuthal angle of the
wavevector $\vec k$ is given by $\phi$. The real part of frequencies
that are solutions of the dispersion relation at a given $\vec k$ are
denoted $\omega_{k\mathrm r}$.

In order to resonate with waves at a given frequency $\omega_{k\mathrm
  r}$ and wavenumber $k_{\parallel}$, particles have to fulfill the
condition
\begin{equation}\label{rescond}
\omega_{k\mathrm r}=k_{\parallel}v_{\parallel}+n\Omega_j
\end{equation}
following from the delta function in Equation~(\ref{qldiff}).  It can be
shown from Equation~(\ref{qldiff}) that alpha particles lose kinetic energy
from resonant wave--particle interactions, thereby acting to drive an
instability,\footnote{For some wave modes, the effect of
  \emph{negative wave energy} has to be taken into account to treat
  the instability correctly. This does not apply to the parallel
  Alfv\'enic instability discussed here. For details about
  negative-energy waves and resonant drift instabilities, we refer to
  the treatment by \citet{verscharen13}.}  if and only if
\begin{equation}
0 < \omega_{k\mathrm r}/k_\parallel < U_\alpha,
\label{eq:condition} 
\end{equation} 
assuming that $f_{\alpha}(\vec v)$ is isotropic about
$U_{\alpha}\hat{\vec e}_z$. A proof of this condition was given in the
Appendix of \citet{verscharen13}.  The requirement that the
drift speed exceed the wave phase velocity along the magnetic field
also arises in the well-studied cosmic-ray streaming instability,
which is excited when the average cosmic-ray drift velocity
along~$\vec{B}_0$ exceeds the Alfv\'en speed~\citep{kulsrud69,wentzel69}.  We
concentrate on the case in which $T_{\perp}=T_{\parallel}$ since we
want to address the effect of the drift on the stability of the A/PC
wave. This drift instability can, however, be assisted or suppressed
by temperature anisotropies as we show in
Section~\ref{hot_plasma}. Protons with $T_{\perp}=T_{\parallel}$
fulfilling the resonance condition will always gain kinetic
energy and, therefore, damp the wave. This is because resonant
interactions cause protons to diffuse in velocity space from regions
of high particle concentration toward regions of low particle
concentration, which means diffusing towards higher energy when
$T_\perp = T_\parallel$.

\section{Analytic Instability Threshold for the Parallel Alfv\'enic Drift Instability}
\label{sect_cold}

In this section, we derive approximate analytic expressions for both
the minimum and maximum values of $U_{\alpha}$ needed for the parallel
Alfv\'enic instability. To simplify the analysis, we use approximate
versions of the dispersion relation (both the cold-plasma dispersion
relation and a non-dispersive approximation). The errors introduced by
these approximations are illustrated in Figures~\ref{fig_disp_omega_cold}
and~\ref{fig_disp_omega}.
\begin{figure}
\epsscale{1.}
\plotone{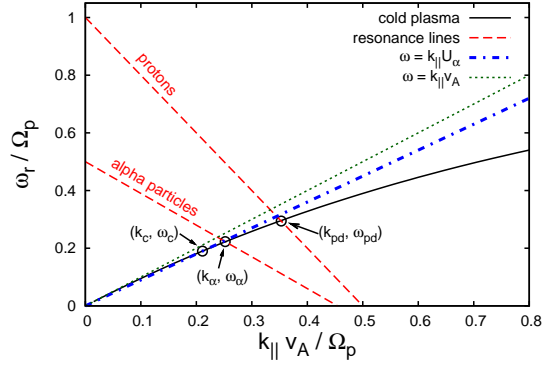}
\caption{The dispersion relation of a cold plasma (black solid line) for $\eta=0.05$, $U_{\alpha}=0.9v_{\mathrm A}$. Red-dashed lines are the resonance conditions Equation~(\ref{rescond}) for protons with $v_{\parallel}=-\Delta v_{\parallel \mathrm p }$ and alpha particles with $v_{\parallel}=U_{\alpha}-\Delta v_{\parallel \alpha}$, where $\Delta v_{\parallel \mathrm p}=\Delta v_{\parallel \alpha}=2v_{\mathrm A}$. The additional lines and labels refer to the definitions in the text.}
\label{fig_cold_situation}
\end{figure}

We begin by determining the critical wavenumber $k_{\mathrm c}$ and
frequency $\omega_{\mathrm c}$ at which the phase speed of the wave
equals $U_{\alpha}$. By combining Equation~(\ref{colddisp}) with the
equation $\omega=k_{\mathrm c}U_{\alpha}$, we find that
\begin{equation}\label{critwave}
\frac{k_{\mathrm c}v_{\mathrm A}}{\Omega_{\mathrm p}}=\frac{1-\left(U_{\alpha}/v_{\mathrm A}\right)^2}{(U_{\alpha}/v_{\mathrm A})}.
\end{equation}
The  frequency at $k_{\parallel}=k_{\mathrm c}$ is simply  $\omega_{\mathrm c}=k_{\mathrm c}U_{\alpha}$.  As discussed in the previous section, $\omega_{k\mathrm r}/k_{\parallel}$ must be less than $U_{\alpha}$ in order for a wave to be driven unstable by resonant interactions with alpha particles when $f_{\alpha}(\vec v)$ is isotropic about $U_{\alpha}\hat{\vec e}_z$. In Figure~\ref{fig_cold_situation}, we plot the cold-plasma dispersion relation, Equation~(\ref{colddisp}), for the parameters $n_{\alpha}=0.05n_{\mathrm p}$ and $U_{\alpha}=0.9v_{\mathrm A}$. As this figure illustrates, $\omega_{k\mathrm r}/k_{\parallel}$ is a monotonically decreasing function of $k_{\parallel}$, which is true in general for the A/PC wave. Because of this, waves with $k_{\parallel}>k_{\mathrm c}$ satisfy the requirement $\omega_{k\mathrm r}/k_{\parallel}<U_{\alpha}$.

We assume now that the distribution functions of the protons and the alpha particles have a finite width $\Delta v_{\parallel \mathrm p }$ and $\Delta v_{\parallel \alpha}$ (defined as positive-definite quantities) in the field-parallel direction in velocity space due to their thermal motion. Therefore, the resonance condition Equation~(\ref{rescond}) can be fulfilled in a range of different particle velocities $v_{\parallel}$ for both species. It is important that the distribution function provides enough particles at the speed $v_{\parallel}$ in Equation~(\ref{rescond}) to drive an instability or to damp the waves efficiently. We assume that  this is the case provided $|v_{\parallel}-U_j| \le 2w_{\parallel j}$. Therefore, we set
\begin{equation}\label{defdeltav}
\Delta v_{\parallel j}=2w_{\parallel j}
\end{equation}
for both protons and alphas.

We assume for concreteness that $\omega_{k\mathrm r} > 0$.  The A/PC
wave with $k_{\perp}=0$ is purely left-handed in polarization with
$E_{kz}=0$ and (since $\omega_{k\mathrm r}>0$) $E_{k,\mathrm
  r}=0$. Equation~(\ref{eq:psink}) then implies that the only relevant
resonance is $n=+1$, because $J_n(0)=\delta_{n,0}$. Only protons that
propagate in the $-z$-direction can participate in the resonant
interaction with the A/PC wave since $\omega_{k\mathrm
  r}<\Omega_{\mathrm p}$ for all $k$ \citep{dusenbery81}. These are
particles with velocities between $v_{\parallel}=0$ and
$v_{\parallel}=-\Delta v_{\parallel \mathrm p}$. The associated proton
damping becomes important for wavenumbers and frequencies larger than
the values $k_{\mathrm{pd}}$ and $\omega_{\mathrm{pd}}$ that are
defined by
\begin{equation}\label{pdwave}
\omega_{\mathrm{pd}}=-k_{\mathrm{pd}}\Delta v_{\parallel \mathrm p}+\Omega_{\mathrm p},
\end{equation}
where $\omega_{\mathrm{pd}}$ is the solution to the dispersion
relation when $k_{\parallel}=k_{\mathrm{pd}}$. The quantities
$\omega_{\mathrm{pd}}$ and $k_{\mathrm{pd}}$ are illustrated in Figure~\ref{fig_cold_situation}.

Because protons are the
majority species, we assume that resonant damping by protons at
$k_{\parallel}>k_{\mathrm{pd}}$ dominates over any possible
instability drive from the alpha particles. Thus, any instabilities
must satisfy $k_{\parallel}<k_{\mathrm{pd}}$. Since unstable modes
must satisfy both $k_{\parallel}>k_{\mathrm c}$ and
$k_{\parallel}<k_{\mathrm{pd}}$, a necessary condition for instability
is that $k_{\mathrm c}<k_{\mathrm{pd}}$. This condition places an
upper limit on $\Delta v_{\parallel \mathrm p }$, which can be obtained
by setting $k_{\mathrm{pd}}>k_{\mathrm c}$ from Equations~(\ref{critwave})
and (\ref{pdwave}):
\begin{equation}\label{first_limit}
\frac{w_{\parallel \mathrm p}}{v_{\mathrm A}}<\frac{1}{2}\left[\frac{\left(U_{\alpha}/v_{\mathrm A}\right)^3}{1-\left(U_{\alpha}/v_{\mathrm A}\right)^2}\right].
\end{equation}
This relation is our first analytic condition for the presence of an
instability. This form is only valid for $U_{\alpha}<v_{\mathrm
  A}$. For larger drift speeds, all solutions of the dispersion
relation have lower phase speeds than $U_{\alpha}$ and fulfill the
instability criterion $\omega_{k\mathrm r}/k_{\parallel}<U_{\alpha}$. Equation~(\ref{first_limit}) can also be interpreted as a lower limit on~$U_\alpha$.

As a second condition for instability, an
appreciable number of alpha particles must satisfy the resonance
condition Equation~(\ref{rescond}) at $k_{\parallel}<k_{\mathrm{pd}}$, so
that alpha particles can drive the instability in the range of
wavenumbers where proton damping is weak. The minimum wavenumber
$k_{\alpha}$ at which thermal alpha particles with
$|v_{\parallel}-U_{\alpha}|\le\Delta v_{\parallel \alpha}$ can satisfy
Equation~(\ref{rescond}) with $n=+1$ is given by the equation
\begin{equation}\label{refcondalpha}
\omega_{\alpha}=k_{\alpha}\left(U_{\alpha}-\Delta v_{\parallel \alpha }\right)+\Omega_{\alpha},
\end{equation}
where $\omega_{\alpha}$ is the solution to the A/PC wave dispersion
relation when $k_{\parallel}=k_{\alpha}$.  For a better understanding
of these labels, we refer again to Figure~\ref{fig_cold_situation}.  We
approximate the dispersion relation now by setting $\omega_{k\mathrm
  r}=k_{\parallel}v_{\mathrm A}$ to solve Equations~(\ref{pdwave}) with
(\ref{refcondalpha}) for $k_{\mathrm{pd}}$ and $k_\alpha$.  The
condition $k_{\alpha}<k_{\mathrm{pd}}$ then yields
\begin{equation}\label{second_limit}
w_{\parallel \alpha  }>\frac{U_{\alpha}+w_{\parallel \mathrm p}}{2}-\frac{1}{4}v_{\mathrm A}.
\end{equation}
This is the second condition for a parallel Alfv\'enic instability.
With the assumption of equal thermal speeds $ w_{\parallel \mathrm p }=
w_{\parallel \alpha }= w_{\parallel}$ for the proton and alpha species,
Equation~(\ref{second_limit}) becomes
\begin{equation}
w_{\parallel}>U_{\alpha}-\frac{1}{2}v_{\mathrm A}.
\end{equation}

Both Equations~(\ref{first_limit}) and (\ref{second_limit}) can be plotted
in the $w_{\parallel \alpha}/v_{\mathrm A}$-$U_{\alpha}/v_{\mathrm A}$ plane to determine parameter
ranges in which the A/PC wave is unstable. This plot is shown
in Figure~\ref{fig_disp_omega_cold}  for the case in which
$w_{\parallel \alpha } = w_{\parallel \mathrm{p}}$.
\begin{figure}
\epsscale{1.}
\plotone{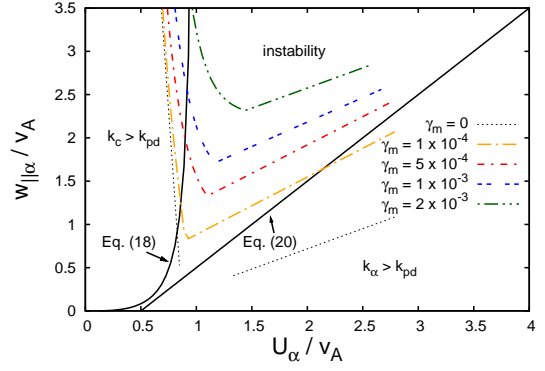}
\caption{Instability criteria in the $w_{\alpha
    \parallel}/v_{\mathrm A}$-$U_{\alpha}/v_{\mathrm A}$ plane under the assumption
  $w_{\parallel \alpha}=w_{\parallel \mathrm p}$. Only the region
  between the two black curves fulfills both approximate analytic
  conditions for instability, Equations~(\ref{first_limit}) and
  (\ref{second_limit}). The contours show maximum growth rates for the
  parallel Alfv\'enic instability calculated by NHDS (see
  Section~\ref{hot_plasma}).}
\label{fig_disp_omega_cold}
\end{figure}

\begin{figure}
\epsscale{1.}
\plotone{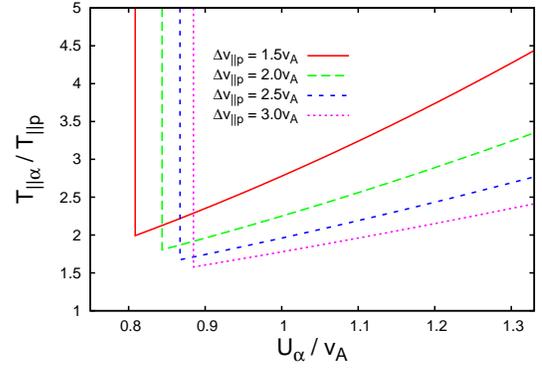}
\caption{Necessary minimum temperature ratio $T_{\parallel \alpha
    }/T_{\parallel \mathrm p}$ for instability. The lines show the thresholds
  for three different values of $\Delta v_{\parallel \mathrm p}$. Only
  temperature ratios above the corresponding line can lead to an
  instability of the parallel Alfv\'enic mode according to
  Equation~(\ref{temp_ratio}). The vertical lines correspond to the minimum $U_\alpha$ values from Equation~(\ref{first_limit}).}
\label{fig_temp_ratio}
\end{figure}

We can also rewrite Equation~(\ref{second_limit}) as a constraint on
the temperature ratio $T_{\parallel \alpha }/T_{\parallel \mathrm
  p}$. Combining Equation~(\ref{second_limit}) with
Equation~(\ref{defdeltav}), we obtain
\begin{equation}\label{temp_ratio}
\frac{T_{\parallel \alpha }}{T_{\parallel \mathrm p}}>4\left[\frac{2U_{\alpha}-v_{\mathrm A}}{2\Delta v_{\parallel \mathrm p}}+\frac{1}{2}\right]^2.
\end{equation}
This condition is plotted in Figure~\ref{fig_temp_ratio}.
It shows that by increasing the values of $\Delta v_{\parallel \mathrm p} $ (i.e., higher parallel proton temperatures) the  minimum value of $T_{\parallel \alpha }/T_{\parallel \mathrm p}$ necessary to excite the parallel Alfv\'enic instability decreases.

\section{Numerical Solutions to the Hot-Plasma Dispersion Relation}\label{hot_plasma}

Using the NHDS code (Section~\ref{dispersions}), we plot in
Figure~\ref{fig_disp_omega_cold} contours of constant maximum growth
rate $\gamma_{\mathrm m}$ from the dispersion relation of a hot plasma consisting of protons, electrons, and alpha particles with  $\eta=0.05$,
$T_{\alpha}=4T_{\mathrm p}=4T_{\mathrm e}$, and $v_{\rm A}/c = 10^{-4}$. All species are assumed to satisfy~$T_\perp = T_\parallel$. The
$\gamma_{\mathrm m}=10^{-4}\Omega_{\mathrm p}$ contour provides a rough match to
our analytic approximation of the instability threshold. This suggests
that the physical interpretation of this instability set forth in the
previous section is approximately valid.

\begin{figure}
\epsscale{1.}
\plotone{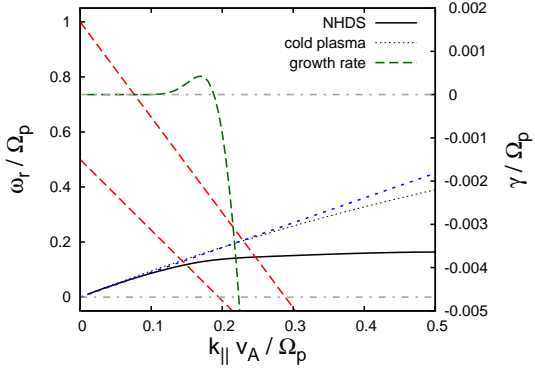}
\caption{Solutions to the dispersion relation of A/PC waves with $\eta=0.05$ and $U_{\alpha}=0.9v_{\mathrm A}$. Solutions in the cold-plasma limit and in a hot plasma with $\beta_{\mathrm p}=3$ are shown. The right vertical axis corresponds to the imaginary part of the NHDS solution.  The blue short-dashed line represents $\omega_{\mathrm r}=k_{\parallel}U_{\alpha}$. The red dashed lines are plots of the resonance conditions for protons and alpha particles for $n=+1$ from Equation~(\ref{rescondline}).}
\label{fig_disp_omega}
\end{figure}

In Figure~\ref{fig_disp_omega}, we illustrate some of the properties of
this instability using numerical solutions to the hot-plasma
dispersion relation  with $\eta=0.05$, $\beta_{\mathrm p}=3$,
$U_{\alpha}=0.9v_{\mathrm A}$, $v_{\mathrm A}/c=10^{-4}$, and
$T_{\alpha}=4T_{\mathrm p}=4T_{\mathrm e}$. 
The cold-plasma solution for $\omega_{k\mathrm r}$ and the NHDS solution for $\omega_{k\mathrm r}$ agree well at low $k_{\parallel}$. At higher $k_{\parallel}$, however, the cold-plasma solution overestimates the frequency of the A/PC wave.
In addition to the two dispersion relations, resonance lines are plotted (red dashed lines) that represent the cyclotron-resonance condition Equation~(\ref{rescond}) for $n=+1$. As in Equations~(\ref{pdwave}) and (\ref{refcondalpha}), these lines are calculated as
\begin{equation}\label{rescondline}
\omega_{\mathrm r}=k_{\parallel}v_{\parallel}+\Omega_j=k_{\parallel}(U_j-2w_{\parallel j})+\Omega_j.
\end{equation}
The parallel speed of the resonant particles again consists of a
component due to the beam (only important for the alpha particles) and
a component due to the thermal width of the particle distribution
function.  The line $\omega_{\mathrm r}=k_{\parallel}U_{\alpha}$ shows
at each wavenumber $k_{\parallel}$ an upper limit for the frequency
$\omega_{k\mathrm r}$ of an unstable wave from the condition
$\omega_{k\mathrm r}/k_{\parallel}<U_{\alpha}$.
Figure~\ref{fig_disp_omega} also shows the linear growth rate $\gamma$
of the A/PC wave calculated by NHDS. As the figure shows, $\gamma>0$
only for $k_{\parallel}$ values between $0.1\Omega_{\mathrm
  p}/v_{\mathrm A}$ and $0.2\Omega_{\mathrm p}/v_{\mathrm A}$. The
endpoints of this interval correspond roughly to $k_{\alpha}$ (and
$k_{\mathrm c}$) and $k_{\mathrm{pd}}$, that is, to the intersections
of the alpha-particle and proton resonance lines with the plot of the
dispersion relation.  At $k_{\parallel}>k_{\mathrm{pd}}$, the damping
by the protons clearly dominates.

The contours of constant $\gamma_{\mathrm m}$ in Figure~\ref{fig_disp_omega_cold}
obtained from the NHDS code can be fit in
$w_{\parallel \alpha }$-$U_{\alpha}$ space. For the nearly vertical portions
of these contours at $U_{\alpha}\lesssim v_{\mathrm A}$, we use a
fitting function of the form
\begin{equation}\label{eqfit}
\frac{w_{\parallel \alpha }}{v_{\mathrm A}}=\left[A+C\left(\frac{U_{\alpha}}{v_{\mathrm A}}\right) ^{\varkappa}\right]^{1/2}.
\end{equation}
The values of the fitting constants for each contour are given in Table \ref{tab_fits}, along with the range of $U_{\alpha}$ values for which the fit applies (defined as the interval between $U_{\min}$ and $U_{\mathrm{med}}$). We fit the shallow-sloped portions of the contours at larger $U_{\alpha}$ with the function
\begin{equation}\label{eqfit2}
\frac{w_{ \parallel \alpha }}{v_{\mathrm A}}=a\frac{U_{\alpha}}{v_{\mathrm A}}+b.
\end{equation}
The best-fit values of $a$ and $b$ are given in Table \ref{tab_fits}, along with the range of $U_{\alpha}$ values, for which the fit is valid (defined as the interval between $U_{\mathrm{med}}$ and $U_{\max}$).

\begin{deluxetable*}{l|ccccc|ccc}
\tablecaption{Fit parameters and limits of the parallel Alfv\'enic instability with isotropic temperatures and $\eta=0.05$. The coefficients for Equation~(\ref{eqfit}) and Equation~(\ref{eqfit2}) are given depending on the maximum growth rate $\gamma_{\mathrm m}$. \label{tab_fits}}
\tablehead{ \colhead{$\gamma_{\mathrm m}/\Omega_{\rm p}$} & \colhead{$A$} & \colhead{$C$} & \colhead{$\varkappa$} &  \colhead{$a$} & \colhead{$b$} & \colhead{$U_{\min}/v_{\mathrm A}$ } & \colhead{$U_{\mathrm{med}}/v_{\mathrm A}$ } & \colhead{$U_{\max}/v_{\mathrm A}$ }  }
\startdata 
$1 \times 10^{-4}$ &-2.0 & 1.3 & -6.59  & 0.66 & 0.22 &0.67 & 0.90 & 2.8  \\
$5\times 10^{-4}$ & 1.0 & 1.45 & -6.79 &  0.66 & 0.61 & 0.68 & 1.09  &2.8  \\ 
$1 \times 10^{-3}$ & 2.0 & 2.55 & -6.52 &  0.57 & 1.05 & 0.74 & 1.17 &2.7  \\
$2\times 10^{-3}$ & 5.0 & 5.45& -7.39 &  0.47 & 1.65 & 0.87 & 1.43 & 2.6
\enddata
\end{deluxetable*}

\begin{figure}
\epsscale{1.}
\plotone{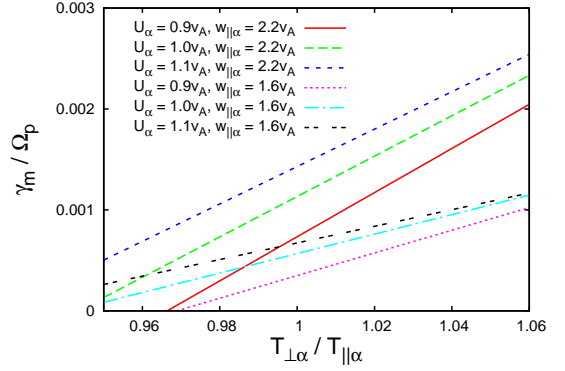}
\caption{The dependence of the maximum growth rate for the parallel Alfv\'enic drift instability on alpha-particle temperature anisotropies. Different combinations of $w_{\parallel \alpha }/v_{\mathrm A}$ and $U_{\alpha}/v_{\mathrm A}$ are shown. In cases with $T_{\perp \alpha }>T_{\parallel \alpha }$, the growth rate is higher than in the isotropic case or at $T_{\perp \alpha }<T_{\parallel \alpha }$. This effect is stronger at higher $w_{\parallel \alpha}/v_{\mathrm A}$.}
\label{fig_anisotropy_dependence}
\end{figure}

High-$\beta_{\alpha}$ plasmas are very sensitive to instabilities driven by temperature anisotropies \citep{gary94,samsonov01,hellinger06,kasper08,bale09,maruca12}. Therefore, we calculate the growth rate of the parallel Alfv\'enic beam instability for different temperature anisotropies of the alpha particles with NHDS. The result of this calculation is shown in Figure~\ref{fig_anisotropy_dependence}.
A small value of the temperature anisotropy can modify the growth rate of the parallel Alfv\'enic drift instability significantly. The growth rate increases with $T_{\perp \alpha }/T_{\parallel \alpha }$. This effect is stronger at higher $w_{\parallel \alpha}/v_{\mathrm A}$ as expected from the behavior of the ion-cyclotron instability without drift \citep{scarf68}. If $T_{\perp \alpha}/T_{\parallel \alpha }>1$, the alpha particles can reach the instability thresholds at lower drift speeds than seen in Figure~\ref{fig_disp_omega_cold}. We note that proton temperature anisotropy can also modify the thresholds of drift instabilities \citep{araneda02,gary03}, but we do not investigate this effect quantitatively in this paper. 

\section{Quasilinear Evolution of the Alpha-Particle Distribution Function}
\label{sec:comp} 

In this section we consider how the alpha particles evolve during resonant
interactions with parallel Alfv\'enic drift instabilities.  
It follows from Equation~(\ref{qldiff}) that resonant wave--particle
interactions cause particles to diffuse in velocity space from regions
of large particle concentration towards regions of smaller particle
concentration.  When particles interact with waves at a single
$k_{\parallel}$ and $\omega_{k\mathrm r}$, the direction of the
diffusive particle flux is constrained to be tangent to semicircles
centered on the parallel phase velocity in the
$v_{\perp}$-$v_{\parallel}$ plane~\citep{kennel66}:
\begin{equation}
v_{\perp}^2+\left(v_{\parallel}-\frac{\omega_{k\mathrm r}}{k_\parallel}\right)^2=\mathrm{constant}.
\end{equation}
These semicircles correspond to curves of constant kinetic energy in
the frame moving at velocity~$(\omega_{k\mathrm r}/k_\parallel)\hat{\vec e}_z$.
When waves are present only 
at a single $\vec{k}$ and~$\omega_{k\mathrm r}$, resonant particles can
only diffuse a tiny distance in $v_{\parallel}$ before falling out of
resonance. However, when a spectrum of waves is present, particles can
undergo velocity-space diffusion over a broader interval of
$v_{\parallel}$. 

\begin{figure}
\epsscale{1.}
\plotone{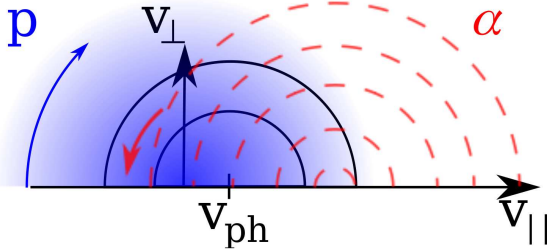}
\caption{Particle densities in velocity space. The distribution function for protons (p, blue) and alpha particles ($\alpha$, red contours) are shown for a high-beta plasma with relative drifts among these species. The solid circles and the red arrow show diffusion directions of the alpha particles. The blue arrow shows the diffusion direction of damping protons. }
\label{fig_diff_paths_highbeta}
\end{figure}

In Figure~\ref{fig_diff_paths_highbeta} we plot the phase space
density of protons and alpha particles for the case in which $U_\alpha
\sim w_{\parallel \alpha } \sim w_{\perp \alpha } \sim
w_{\parallel \mathrm{p}} \sim w_{\perp \mathrm{p}}$.  As illustrated in
Figures~\ref{fig_cold_situation} and~\ref{fig_disp_omega}, the alpha
particles that resonate with the parallel Alfv\'enic drift instability
typically satisfy $v_\parallel < 0$ and thus reside on the left side
of Figure~\ref{fig_diff_paths_highbeta}.  When these particles diffuse
from interactions with instabilities at some $\vec{k}$ and $\omega_{k\mathrm r}$,
their diffusive flux is parallel to the semicircular contours centered
on the point $(v_{\mathrm{ ph}}, 0)$, where $v_{\mathrm{ph}} \equiv
\omega_{k\mathrm r}/k_\parallel$. Because they diffuse from high particle
concentration to low particle concentration, they migrate in the
direction of the red arrow, towards smaller~$v_\perp$. This diffusion
thus acts to ``pinch'' the alpha-particle distribution at $v_\parallel
< 0$, which leads to the appearance of a parallel-beam-like feature
(narrow in $v_\perp$ but broader in $v_\parallel$) propagating in the
$-z$ direction. This type of feature has been found in measurements of
the alpha-particle distribution function in the solar wind at
high $\beta_{\alpha}$ \citep[see for example Figure~4
  in][]{astudillo96}.

\section{Comparison with Observations}\label{sect:obs}

\begin{figure}
\epsscale{1.}
\plotone{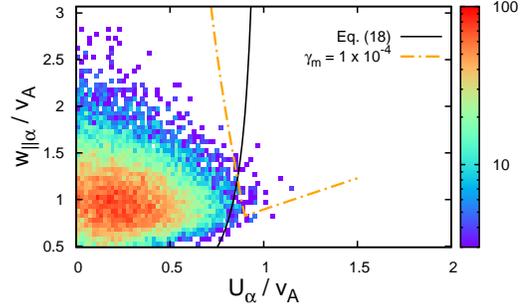}
\caption{Comparison of theoretical instability thresholds and measurements from the \emph{Wind} spacecraft. The color coding represents the number of spectra per bin of the measurement. We only show observations in the parameter range relevant to the parallel Alfv\'enic instability with $A_{\mathrm c}\le 0.3 $, $0.9\le T_{\perp \alpha}/T_{\parallel \alpha}\le 1.1$, and $3.5\le T_{\parallel \alpha}/T_{\parallel \mathrm p}\le 4.5$. The black solid line shows the analytical threshold from Equation~(\ref{first_limit}) for $w_{\parallel \mathrm p}=w_{\parallel \alpha}$, and the orange line shows an isocontour of constant maximum growth rate obtained from the NHDS code for isotropic alpha and proton temperatures.}
\label{fig_Uw_data}
\end{figure}

To investigate the possible relevance of the parallel Alfv\'enic instability to alpha particles in the solar wind, we consider measurements of alpha-particle differential flow from the \emph{Wind} spacecraft at 1 AU. 
We use ion measurements from \emph{Wind}'s two Faraday cups, which are
part of the spacecraft's Solar Wind Experiment \citep{ogilvie95}. 
The cups together deliver one ion spectrum about every 90 s. 
Several versions of automated code have been developed for deriving values
for the bulk parameters (i.e., density, velocity, and temperature) of
protons and alpha particles from each spectrum.  For this study, we use
the output from the \citet{maruca12b} code, which incorporates
3 s magnetic field measurements from \emph{Wind}'s Magnetic Field
Investigation \citep{lepping95} into its nonlinear fitting of each
ion spectrum in order to separate the perpendicular and parallel
components of velocity and temperature.  This code has processed all
$4.8\times10^6$ \emph{Wind} ion spectra from late-1994 (i.e., the
spacecraft's launch) through mid-2010.  After removing spectra with poor
signal, with a high collisional age \citep{kasper08}, and/or from near
or within the Earth's bow shock, $9.3\times10^5$ spectra remain
\citep{maruca12b}.  We further reduce this dataset to the
$3.0\times10^4$ spectra that also satisfy $0.9 \le T_{\perp \alpha}/T_{\parallel \alpha} \le 1.1$ and
$3.5 \le T_{\parallel \alpha}/T_{\parallel \mathrm{p}} \le 4.5$.
 Figure~\ref{fig_Uw_data} shows the distribution of data  in the same plane that we use in Figure~\ref{fig_disp_omega_cold}. We also plot in Figure~\ref{fig_Uw_data} the instability threshold.
The data distribution reaches the instability threshold in the range $0.7 \lesssim w_{\parallel \alpha}/v_{\mathrm A}\lesssim 1.5$. At lower thermal speeds, other instabilities such as the parallel magnetosonic instability have lower thresholds.

It is not entirely clear what to conclude from the fact that the instability threshold approximately bounds the data distribution in Figure~\ref{fig_Uw_data}. This figure suggests that the parallel Alfv\'enic instability acts as a deceleration mechanism that limits $U_{\alpha}$ to values below the instability threshold. However, a problem with this scenario is that, as discussed in Section~\ref{sec:comp}, the parallel Alfv\'enic instability resonantly interacts only with alpha particles with $v_{\parallel}<0$ in the proton rest frame. Although the ``pinching'' effect described in Section~\ref{sec:comp} leads to some reduction of $U_{\alpha}$, the bulk of the alpha particles, which satisfy $v_{\parallel}>0$, are unaffected by the instability. Thus, although the parallel Alfv\'enic instability may contribute to alpha-particle deceleration, it is unable on its own to explain how alpha particles decelerate between heliocentric distances of 0.3 AU and 1 AU. 

 If the solar wind were to expand according to the double-adiabatic prediction \citep{chew56}, then alpha particles would satisfy $T_{\perp\alpha}\ll T_{\parallel \alpha}$ at 1 AU. Under this condition, the parallel Alfv\'enic instability is not unstable at the observed values of $U_{\alpha}$. 
However, spacecraft measurements show that there are solar-wind streams at 1 AU with $T_{\perp \alpha}\simeq T_{\parallel \alpha}$. In these cases, the alpha particles have undergone perpendicular heating, perhaps from the dissipation of solar-wind turbulence, and/or the excitation of magnetosonic instabilities, whose nonlinear evolution (unlike that of the parallel Alfv\'enic instability) leads to an increase in $T_{\perp \alpha}/T_{\parallel \alpha}$ \citep{gary00b}. Thus, although Figure~\ref{fig_Uw_data} shows that the parallel Alfv\'enic instability occurs in the solar wind, this instability does not offer a complete explanation of the radial evolution of alpha-particle properties in the solar wind.

\section{Discussion and Conclusions}\label{conclusions}

Using results from quasilinear theory, we derive approximate analytic expressions describing the conditions under which the parallel-propagating A/PC wave becomes unstable in the presence of alpha particles drifting parallel to the magnetic field at speed $U_{\alpha}$. We assume that $T_{\perp \alpha} = T_{\parallel \alpha}$. To obtain these expressions, we consider the competing effects of the instability drive provided by the alpha particles and the cyclotron damping caused by thermal protons. We then find that there are two conditions needed for this instability to arise. First, $U_{\alpha}$ must be sufficiently large that there are waves with $\omega_{k\mathrm r}/k_{\parallel} < U_{\alpha}$ at wavenumbers that are sufficiently small that proton cyclotron damping can be neglected. Second, the alpha-particle thermal speed must be sufficiently large that the alpha particles can resonate with the wave at wavenumbers that are too small for thermal-proton cyclotron damping to occur. These two conditions lead to Equations~(\ref{first_limit}) and~(\ref{second_limit}), respectively.
Both resonant alpha particles and resonant protons can only fulfill
these conditions if they have $v_{\parallel}<0$ in the proton frame.
A comparison with solutions from the full dispersion relation of a hot
plasma in Section~\ref{hot_plasma} shows rough agreement with our
analytical expressions for the instability thresholds in
Equations~(\ref{first_limit}) and (\ref{second_limit}). 

In the fast solar wind, $T_{\parallel \alpha  }$ is typically $\simeq 4
T_{\parallel \mathrm{p}}$ and $\eta \simeq
0.05$~\citep{bame77,kasper08}. Under these conditions, we find that the
minimum $U_\alpha$ thresholds for the parallel Alfv\'enic drift
instability are in the range of $0.7v_{\mathrm A}$ to $0.9v_{\mathrm
  A}$ for $1\lesssim \beta_{\mathrm p}\lesssim 12$.  These $U_\alpha$
thresholds are comparable to the limits on $U_\alpha$ that are seen in
the solar wind, suggesting that this instability may be important for
limiting alpha-particle differential flow in the solar wind when
$\beta_{\rm p}\gtrsim 1$.

Although we have focused on the case in which $T_{\perp } = T_{\parallel }$
for all particle species, we find that the growth rate of the parallel
Alfv\'enic drift instability increases with increasing~$T_{\perp \alpha}/T_{\parallel \alpha }$ and that the parallel Alfv\'enic
instability is more strongly affected by temperature anisotropy when
$\beta_{\parallel \alpha}$ is higher. These trends are also seen
in the case of the ion-cyclotron instability driven by
alpha-particle temperature anisotropy in the absence of differential
flow~\citep{maruca12}.

The parallel Alfv\'enic instability alone does not offer a full description of the alpha-particle evolution in the solar wind. Local perpendicular heating and additional instabilities are necessary in order to explain the properties of the alpha particles. 
Measurements by the \emph{Wind} spacecraft presented in Section~\ref{sect:obs}, however, indicate that the thresholds of the parallel Alfv\'enic instability are reached in some solar wind streams with $w_{\parallel \alpha}/v_{\mathrm A}\gtrsim 0.7$ at 1 AU. In this range and for $T_{\perp \alpha}= T_{\parallel\alpha}$, the discussed instability has the lowest threshold of the known linear instabilities. We do not treat the mechanisms that lead to the observed conditions in detail. Nevertheless, in these solar-wind streams, the parallel Alfv\'enic instability contributes to the alpha-particle evolution and generates parallel-propagating A/PC waves. At lower $w_{\parallel \alpha}/v_{\mathrm A}$, oblique Alfv\'en/ion-cyclotron instabilities are likely able to regulate the drift efficiently \citep{gary00,li00,verscharen13}. 

In closing, we note that the parallel Alfv\'enic drift instability is similar to the cosmic-ray
streaming instability~\citep{kulsrud69,wentzel69}. Both instabilities
require that the drifting ion population have an average velocity
along~$\vec{B}_0$ that exceeds $\omega_{k\mathrm r}/k_{\parallel}$. The principal
differences between the two instabilities are that the parallel
Alfve\'nic drift instability involves thermal particles, dispersive
waves, and a competition between the instability drive of the
streaming ion population and the resonant cyclotron damping by
thermal protons. Because of these differences, the instability
criteria are different in the two cases.
Another similar instability is the ion/ion left-hand resonant instability \citep{gary93}---also known as the ion-ion L-mode instability \citep{treumann97}---in which the parallel Alfv\'en/ion-cyclotron wave is driven unstable by resonant particles at $v_{\parallel}<0$ when a hot beam is present with a bulk speed $U_{\mathrm b}>0$ on the tail of the main-ion species. The instability that we investigate differs from the ion-ion L-mode instability in that we include a second ion species. Also, our analysis goes beyond previous investigations of the ion-ion L-mode instability by examining the competition between the destabilizing effects of the beam and the resonant cyclotron damping by the core particles, and by calculating analytical and numerical instability thresholds.

\acknowledgements

We appreciate helpful discussions with Stuart Bale, Eliot Quataert, Joe Hollweg, Marty Lee, and
Phil Isenberg.  We thank Justin Kasper for his assistance in analyzing the \emph{Wind} Faraday cup data. This work was supported in part by grant NNX11AJ37G
from NASA's Heliophysics Theory Program, NASA grant NNX12AB27G,
NSF/DOE grant AGS-1003451, and DOE grant DE-FG02-07-ER46372.

\bibliographystyle{apj}
\bibliography{parallel_alfvenic}
\end{document}